\documentclass[iop]{emulateapj}
\usepackage{natbib} 
\def\lapp{\ifmmode\stackrel{<}{_{\sim}}\else$\stackrel{<}{_{\sim}}$\fi}
\def\gapp{\ifmmode\stackrel{>}{_{\sim}}\else$\stackrel{>}{_{\sim}}$\fi}
\usepackage{multirow}
\usepackage{color}
\usepackage{subfigure}
\usepackage{bm}
\usepackage{lmodern}
\usepackage{soul}
\usepackage{amsmath}

\begin{document}

\title{An Improved Transit Measurement for a 2.4 $R_{\Earth}$ Planet Orbiting A Bright Mid-M Dwarf K2$-$28} 

\author{Ge Chen\altaffilmark{1}, Heather A. Knutson\altaffilmark{2}, Courtney D. Dressing\altaffilmark{2,3,14}, Caroline V. Morley\altaffilmark{4, 14}, Michael Werner\altaffilmark{5}, Varoujan Gorjian\altaffilmark{5}, Charles Beichman\altaffilmark{6}, Bj\"{o}rn Benneke\altaffilmark{7}, Jessie Christiansen\altaffilmark{6}, David Ciardi\altaffilmark{8}, Ian Crossfield\altaffilmark{9}, Steve B. Howell\altaffilmark{10}, Jessica E. Krick\altaffilmark{8}, John Livingston\altaffilmark{11}, Farisa Y. Morales\altaffilmark{5, 12}, Joshua E. Schlieder\altaffilmark{13} 
 }
\affil{
{\small $^1$ Department of Physics, California Institute of Technology, Pasadena, CA 91125, USA;  gcchen@caltech.edu}\\ 
{\small $^2$ Division of Geological \& Planetary Sciences, California Institute of Technology, Pasadena, CA 91125, USA}\\ 
{\small $^3$ Department of Astronomy, University of California at Berkeley, Berkeley, CA 94720, USA}\\ 
{\small $^4$ Harvard University, Center for Astrophysics, 60 Garden St. Cambridge MA 02138, USA}\\ 
{\small $^5$ Jet Propulsion Laboratory, California Institute of Technology, 4800 Oak Grove Drive, Pasadena, CA 91109, USA}\\
{\small $^6$ NASA Exoplanet Science Institute, California Institute of Technology, Jet Propulsion Laboratory, Pasadena, CA 91125, USA }\\
{\small $^7$ D\'{e}partement de Physique, Universit\'{e} de Montr\'{e}al, Montreal, H3T 1J4, Canada}\\ 
{\small $^8$ IPAC MC 314-6 California Institute of Technology 1200 E. California Blvd. Pasadena, CA, 91125, USA}\\
{\small $^9$ MIT-Kavli Institute for Astrophysics and Space Research, 70 Vassar Street, Cambridge, MA 02139, USA}\\
{\small $^{10}$ Space Science \& Astrobiology Division, NASA Ames Research Center, Moffett Field, CA 94035, UAS}\\
{\small $^{11}$ Department of Astronomy, The University of Tokyo, 7-3-1 Hongo, Bunkyo-ku, Tokyo 113-0033, Japan}\\
{\small $^{12}$ Department of Physical Sciences, Moorpark College,7075 Campus Road, Moorpark, CA 93021, USA}\\
{\small $^{13}$ Exoplanets and Stellar Astrophysics Laboratory, Code 667, NASA Goddard Space Flight Center, Greenbelt, MD 20771, USA}\\
}

\altaffiltext{14}{NASA Sagan Fellow}

\begin{abstract}
We present a new {\em Spitzer} transit observation of K2$-$28b, a sub-Neptune ($R_{\rm p} = 2.45\pm0.28 R_{\Earth}$) orbiting a relatively bright ($V_{\rm mag} = 16.06$, $K_{\rm mag} = 10.75$) metal-rich M4 dwarf (EPIC 206318379). This star is one of only seven with masses less than 0.2 M$_{\Sun}$ known to host transiting planets, and the planet appears to be a slightly smaller analogue of GJ 1214b ($2.85\pm0.20 R_{\Earth}$; \citealt{2013A&A...549A..10H}). 
Our new {\em Spitzer} observations were taken two years after the original K2 discovery data and have a significantly higher cadence, allowing us to derive improved estimates for this planet's radius, semi-major axis, and orbital period, which greatly reduce the uncertainty in the prediction of near future transit times for the {\em James Webb Space Telescope} ({\em JWST}) observations. We also evaluate the system's suitability for atmospheric characterization with {\em JWST} and find that it is currently the only small ($< 3 R_{\Earth}$) and cool ($< 600$ K) planet aside from GJ 1214b with a potentially detectable secondary eclipse.  We also note that this system is a favorable target for near-infrared radial velocity instruments on larger telescopes (e.g., the Habitable Planet Finder on the Hobby-Eberly Telescope), making it one of only a handful of small, cool planets accessible with this technique. Finally, we compare our results with the simulated catalog of the Transiting Exoplanet Survey Satellite ({\em TESS}) and find K2$-$28b to be representative of the kind of mid-M systems that should be detectible in the {\em TESS} sample. 
\end{abstract}


\section{Introduction} 
\label{sec:intro} 
In the four years since the end of the original Kepler mission, the NASA {\em K2} mission (\citealt{2014PASP..126..398H}) has sought to increase the known number of exoplanets orbiting nearby low-mass stars. M dwarfs hosts, especially mid to late M-dwarfs, are particularly advantageous for detailed characterization studies as the small stellar radius, low stellar mass and low stellar temperature results in a larger transmission signal, secondary eclipse depth, and radial velocity for a given planet (\citealt{2008PASP..120..317N}, \citealt{2014prpl.conf..739M}, \citealt{2015ApJ...807...45D}). 
Although these late type stars are typically too faint for most optical radial velocity instruments, they are accessible at red optical (\citealt{2016A&A...586A.101F}) and infrared wavelengths (e.g., \citealt{2012SPIE.8446E..1SM}, \citealt{2012SPIE.8446E..0RQ}, \citealt{2013A&A...557A.139S}) and exhibit relatively large radial velocity signals due to their smaller stellar masses. As demonstrated by the interest in systems like GJ1214b and TRAPPIST-1 (\citealt{Charbonneau:2009aa}; \citealt{2016ApJ...829L...2H}; \citealt{Gillon:2017aa}), mid to late M dwarfs remain a key touchstone in our quest to understand the nature and origin of planetary systems.

In this study we focus on K2$-$28b, a sub-Neptune-sized planet orbiting the M4 dwarf K2$-$28 (EPIC 206318379). The planet was discovered in {\em K2} Campaign 3 by \citet{2016ApJS..222...14V} 
and statistically validated by \citet{2016ApJ...820...41H}. The host star has a mass of $0.20^{+0.09}_{-0.10}$ M$_{\Sun}$, a radius of $0.28\pm0.03$ R$_{\Sun}$, and an effective temperature of  $3290\pm90$ K (\citealt{2017ApJ...836..167D}).  It is one of only seven stars with masses less than or equal to $0.2 M_{\Sun}$ known to host transiting planets (16 planets in total), and with $\rm K_{mag} = 10.75$ it is the fifth brightest of these systems. \citet{2016ApJ...820...41H} reported a planet radius of $2.23\pm0.25$ R$_{\Earth}$, an orbital period of 2.26 days, and an equilibrium temperature of $\sim$500 K, based on a joint analysis of the {\em K2} data as well as ground-based follow-up transit observations made with the Simultaneous Infrared Imager for Unbiased Survey (SIRIUS; \citealt{2003SPIE.4841..459N}) on the Infrared Survey Facility 1.4m telescope and the Multicolor Simultaneous Camera for studying Atmospheres of
Transiting exoplanets (MuSCAT; \citealt{2015JATIS...1d5001N}) on the Okayama 1.88 m telescope. 
More recently, \citet{2017ApJ...836..167D} presented refined stellar properties for a sample of low-mass K2 stars including K2$-$28 using empirical relations from \citet{0004-637X-800-2-85}, resulting in an updated planet radius estimate of $2.30\pm0.27$ R$_{\Earth}$ (\citealt{2017AJ....154..207D}). 
However, the 30 minute cadence of the {\em K2} data is poorly matched to its relatively short one hour transit duration, and the current ground-based transit observations from \citet{2016ApJ...820...41H} have a relatively low signal-to-noise ratio, making it difficult to characterize the transit shape accurately. 

We obtained a {\em Spitzer} transit observation of K2$-$28b as part of an ongoing large program focused on follow-up of new transiting planets detected by the {\em K2} mission (GO 13052, PI Werner; this dataset ID is ADS/Sa.Spitzer\#62339840). Thanks to its short cadence and long baseline, this {\em Spitzer} observation can in principle significantly improve the planetary characterization and orbital ephemerides as compared to fits using the K2 data alone (\citealt{2016ApJ...822...39B}). An accurate ephemeris is particularly important for atmospheric characterization studies with space telescopes, as large uncertainties in the predicted time of transit can significantly increase the overheads associated with these observations (\citealt{2017ApJ...834..187B}).
In this paper we carry out a joint fit to the K2$-$28b transit data observed by {\em Spitzer} and {\em K2} in order to obtain improved parameters for this system and evaluate its suitability for future observations with the {\em James Webb Space Telescope} ({\em JWST}), the Transiting Exoplanet Survey Satellite ({\em TESS}), and near-infrared radial velocity instruments. 

In Section \ref{sec:obs}, we describe the observations. In Section \ref{sec:ana}, we explain the data reduction methods and present our results. In Section \ref{sec:disc}, we discuss the implications of these results for our understanding of this system as well as M dwarf planetary systems in general, and we conclude in Section \ref{sec:con}.

\section{Observations and Photometry} 
\label{sec:obs} 
\subsection{{\em Spitzer} 4.5 $\mu$m Photometry} 
\label{sec:spitzerobs}
K2$-$28 was observed by {\em Spitzer} on UT 2017 Febuary 11 for a total of 7.9 hours. The observation was made using the 4.5 $\mu$m bandpass of the Infra-Red Array Camera (IRAC; \citealt{2004ApJS..154...10F}).  Images were obtained in subarray mode using the standard peak-up pointing method in order to improve the initial placement of the star on the array (\citealt{2012SPIE.8442E..1YI}).
The data consist of 14,208 frames with an exposure time of 2 s for each image.  

\begin{figure}[] 
  \centering
  \includegraphics[width=0.5\textwidth]{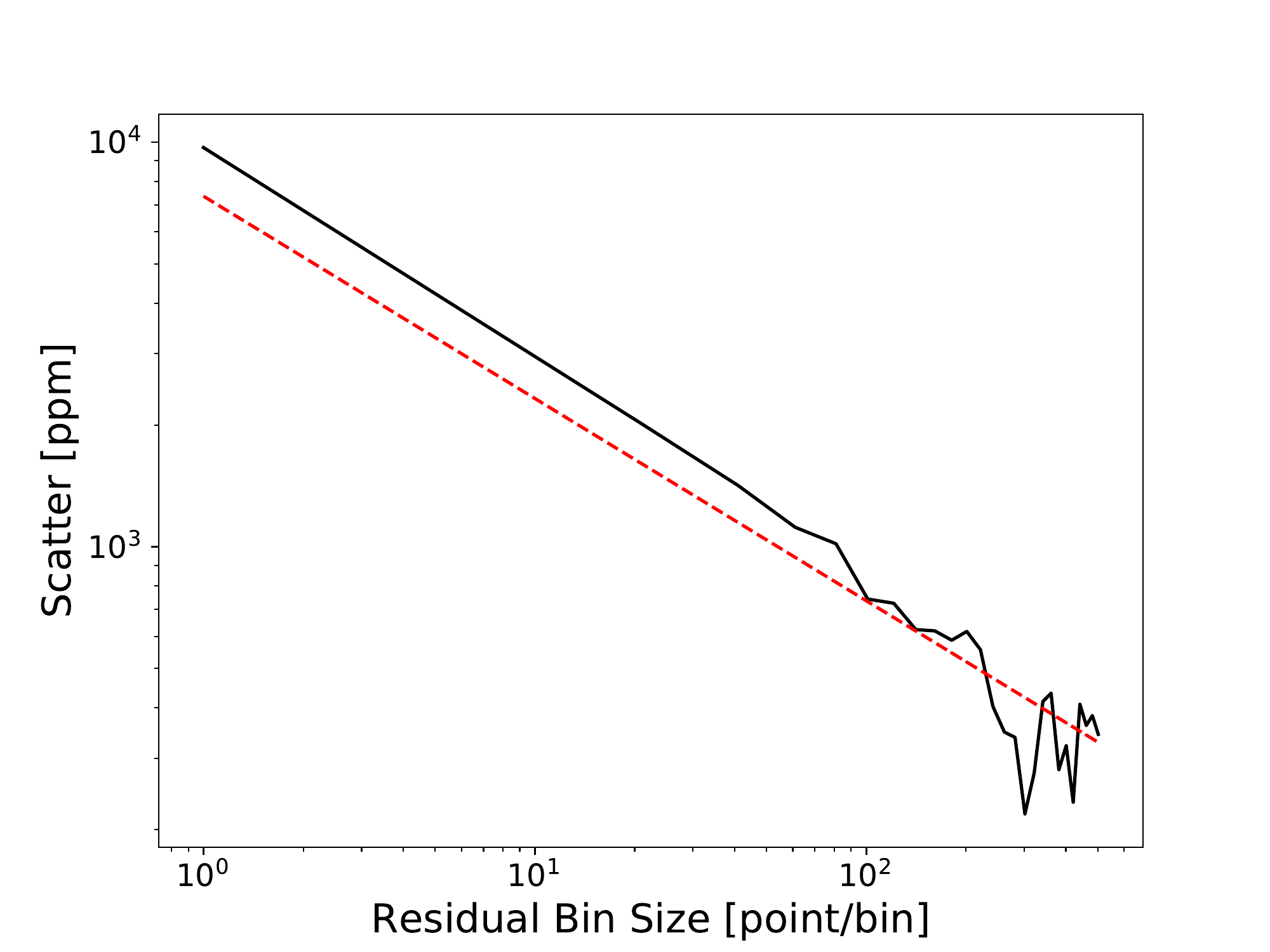}
  \caption{RMS of the {\em Spitzer} photometry versus bin size. The solid black line shows the standard deviation of the residuals from our best-fit solution as a function of bin size, and the dashed red line is the predicted photon noise scaled as the square root of the number of points in each bin. The scatter in the data is 1.25 times the photon noise limit for the unbinned light curve, but decreases to a value closer to the photon noise limit at the largest bin sizes.
\label{fig:scatter}
}
\end{figure}

The {\em Spitzer} photometry was extracted using a circular aperture following the methods described in \citet{2017arXiv171007642Z}.  
We first estimated the sky background by calculating the mean location of the fluxes in pixels located more than twelve pixels away from the location of the star, and subtracted this background from each $32\times32$ pixel image.  We then used flux-weighted centroiding to calculate the position of the star in each image, and centered our photometric aperture at that position. We considered 20 radii for our circular photometric aperture ranging between 1.5 to 5.0 pixels. We also binned the data before fitting, as previous studies have shown that this minimizes the noise on longer timescales relevant for the transit (e.g., \citealt{2015ApJ...805..132D}).  We considered bin sizes of 10, 20, 30, 40, 50, 60, 90 and 120 points per bin (i.e., 20-240 seconds). 
We determined the optimal choices for the aperture radius and bin size by calculating the squared distance between the standard deviation of the residuals from our best-fit solution versus bin size and the predicted photon noise scaled as the square root scale (Fig. \ref{fig:scatter}). We found that we preferred a radius of 1.9 pixels and a bin size of 40 points (80 seconds) per bin. After choosing the aperture radius and bin size, we followed the same criteria and found that we obtained the lowest scatter across all timescales when we trimmed the first and last 2.1 hours of our observations. 

Although this is a relatively large trim duration, we found that these observations exhibited a time-varying trend with an approximately quadratic shape, which could be due either to uncorrected instrumental noise or to stellar activity.  Rather than fitting a quadratic function to our data, which can affect our estimates of the transit depth (e.g., \citealt{2010ApJ...721.1861A}), 
we instead opted to keep our original linear function and fit over a smaller segment of data.  Our final trimmed light curve has a total duration of 3.7 hours, which is still significantly longer than the planet's one-hour transit duration. We see no evidence for correlated noise in the data on these timescales, and achieve a scatter of 1.25 times the predicted photon noise limit.

\subsection{K2 Photometry} 
\label{sec:k2obs}
We also re-analyzed the previously reported optical {\em K2} data, which has an integration time of 30 minutes and was obtained between UT 2014 November 15 and UT 2015 January 23 (\citealt{2016ApJS..222...14V}).  The total duration of this observation was 69 days, spanning 29 orbits of the planet.  We utilize the version of the {\em K2} K2SFF photometry provided by Andrew Vanderburg (\citealt{2016ApJS..222...14V}), which includes a correction for short-term spacecraft systematics fit simultaneously with the transits. The remaining structure in the photometry is very likely due to long-term instrumental systematics. We trimmed segments of this light curve centered on individual transit events, and considered baselines ranging between $1.9-7.7$ hours of data before the ingress and after the egress of each event.  
We found that a baseline of 3.4 hours on either side minimized the scatter in the residuals from our fit to the phased light curve.

\section{Light Curve Analysis and Results} 
\label{sec:ana} 
\subsection{Spitzer Light Curve} 
\label{sec:spitzerlcv} 
\begin{figure*} 
  \centering
  \subfigure[]{\label{fig:spitzer}\includegraphics[width=0.49\textwidth]{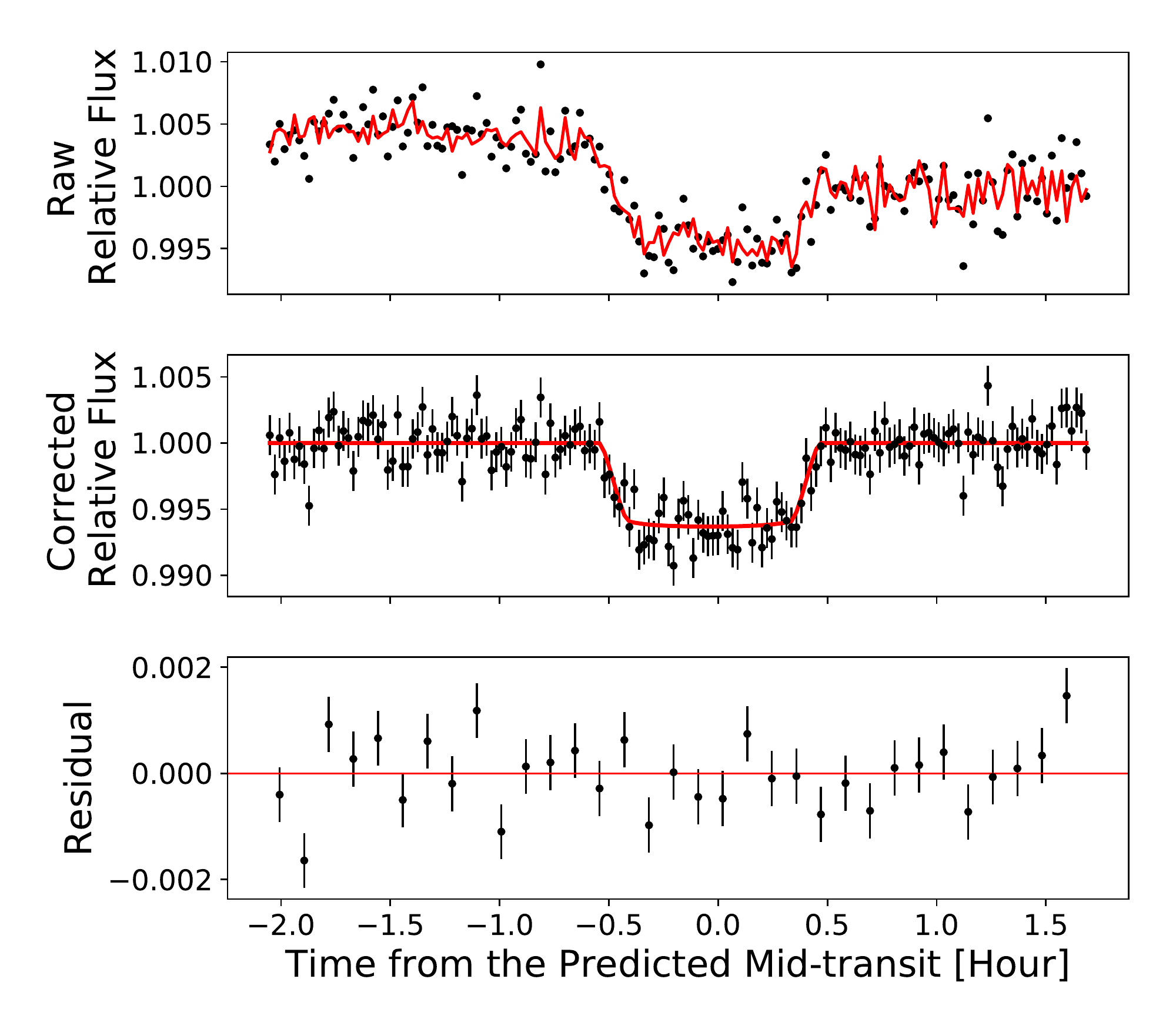}}
  \subfigure[]{\label{fig:k2}\includegraphics[width=0.49\textwidth]{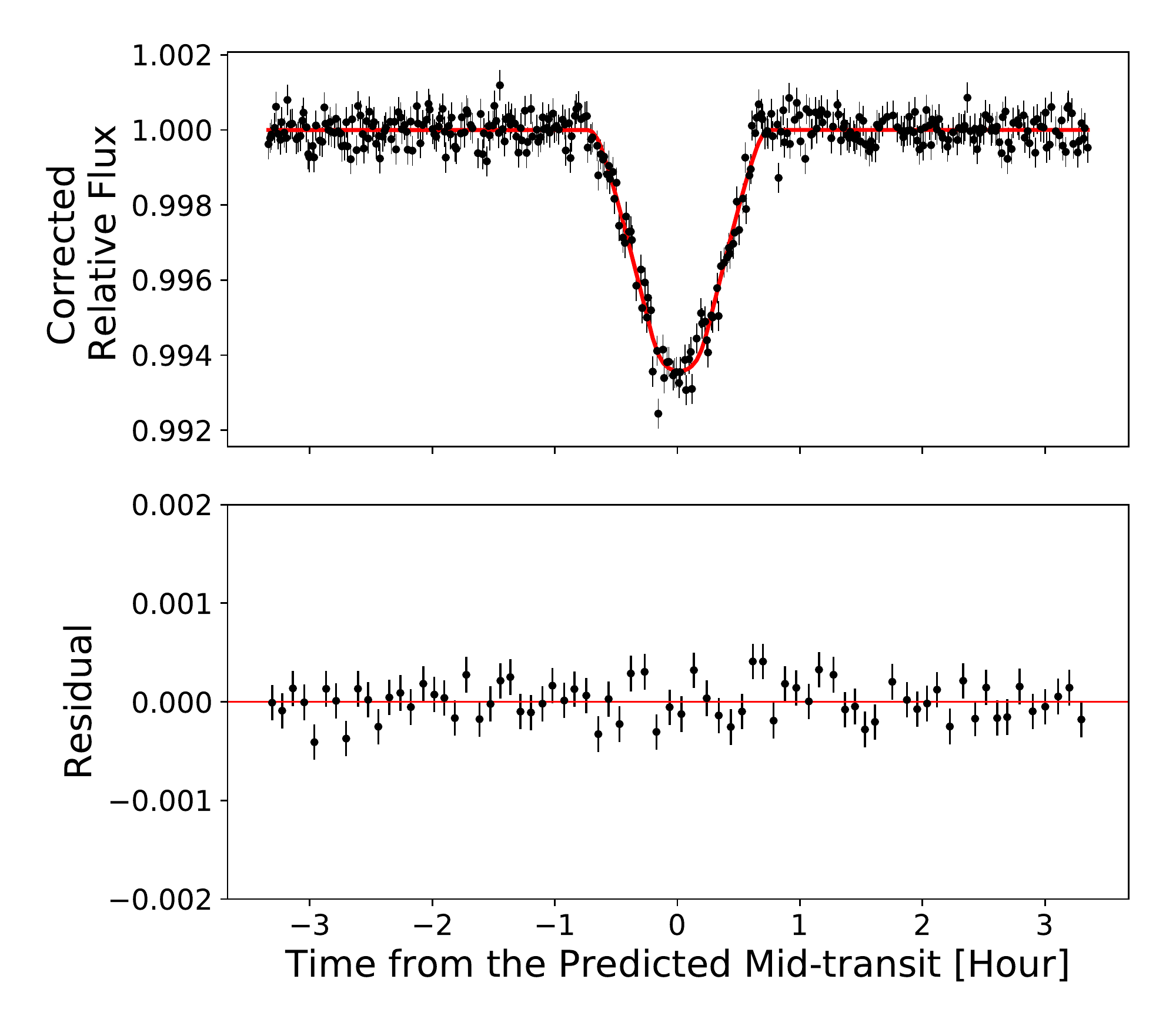}}
  \caption{Best-fit transit light curves calculated from a joint fit to both data sets (Section \ref{sec:jointlcv}). (a) 4.5 $\mu$m {\em Spitzer} observation obtained on UT 2017 Feb 11, shown with a bin size of 40 ($\sim$80 seconds). 
 Top panel: normalized {\em Spitzer} raw photometry (grey filled circles) and best-fit model including both astrophysical and instrumental components (red line). Middle panel: {\em Spitzer} relative flux (grey filled circles) and the best-fit transit light curve model (red line) after dividing out the instrumental model, with uncertainties set to the noise parameter $\sigma_{\rm Spitzer}$. Bottom: Fit residuals (data $-$ model) binned to 200 points (400 seconds) per bin for display. (b) Phase-folded {\em K2} transit observations obtained between UT 2014 Nov 15 and UT 2015 Jan 23; the increased curvature in the transit shape relative to the {\em Spitzer} observations is due to longer (30 minute versus 2 s) integration times, along with more prominent limb-darkening at optical wavelengths. Top panel: systematic-corrected data (grey filled circles) and the best-fit transit light curve model (red), with uncertainties set to the noise parameter $\sigma_{\rm K2}$. Bottom panel: Fit residuals (data $-$ model) binned to 5 points per bin for display. 
\label{fig:lcv}
}
\end{figure*}

{\em Spitzer} IRAC data exhibit systematic errors due to intra-pixel sensitivity variations combined with telescope pointing variations (e.g., \citealt{0004-637X-626-1-523}). We corrected for this effect using the pixel-level decorrelation (PLD) method described in \citet{2015ApJ...805..132D}.  In this study we utilized an updated version of this model (\citealt{2017ApJ...834..187B}), which treats this as a multiplicative rather than an additive effect and removes the constant term: 
\begin{equation}
\label{eq:instru_s} 
S(t_i) = \frac{\sum_{k=1}^{9}w_{k}P_{k}(t_i)}{\sum_{k=1}^{9}P_k(t_i)}+m\cdot t_i. 
\end{equation}
Here, $S(t_i)$ is the flux measured  
in the $i^{\rm th}$ image, $t_i$ is the time from predicted transit center, $P_k$ is the flux in each pixel of a $3\times 3$-pixel square array centered on the star, $w_k$ sets the relative weight of each pixel in the model, and $m$ is a slope that accounts for temporal variations. Although previous studies typically included the nine $w_k$ variables as free parameters in the fit, we instead determined their values at each step of our fit using linear regression.  We found that this allowed our fits to converge more quickly and reliably to the correct solution, as these nine parameters were otherwise highly degenerate.  This approach also has the added advantage of reducing the number of free parameters in our model, which is helpful for the global fit including the {\em K2} data. We noted that minimizing these parameters using linear regression can lead us to underestimate the uncertainties in our best-fit transit parameters.  We tested the magnitude of this effect by carrying out a second version of the fits in which these weights were included as free parameters.  We found that our final uncertainties on the transit parameters for the {\em Spitzer} data varied by less than 5\% in this version, indicating that these weights did not contribute significantly to the uncertainties in the transit shape parameters. 

We calculated the transit light curve model using the {\ttfamily Batman} package (\citealt{2015PASP..127.1161K}), where we assumed a circular orbit for the planet and fixed the period to the value described in \citet{2016ApJ...820...41H}. We allowed the planet-star radius ratio $R_{\rm p}/R_{\rm *}$, impact parameter $b$, semi-major axis ratio $a/R_{\rm *}$, and center of transit time $T_{0}$ and the scatter $\sigma_{\rm Spitzer}$ to vary as free parameters. 
We fixed the quadratic limb darkening coefficients ($u_1=0.0092, u_2=0.1877$) to band-integrated values based on the Atlas stellar models from \citet{2010A&A...510A..21S}. We selected limb-darkening parameters for a star with $T_{\rm eff}=3500$ K, log($g$) = 5.00, and [M/H] = 0.20, which are in reasonably good agreement with the published parameters for K2$-$28 ($T_{\rm eff} = 3290\pm90$ K, log($g$) = $4.93\pm0.24$, and [M/H] = $0.208\pm0.090$ \citealt{2017ApJ...836..167D}). We also re-ran our fits with limb-darkening coefficients for models $250$ K hotter and log($g$) 0.5 lower than our nominal model, and with limb-darkening coefficients calculated based on 1D Phoenix stellar model  (\citealt{2012A&A...546A..14C}) for a star with $T_{\rm eff}=3300$ K and log($g$) = 5.00, and found that our resulting transit shape parameters changed by less than 0.1$\sigma$. Although \citet{2010A&A...510A..21S} only provided limb darkening coefficients for stars hotter than $3500$ K, we conclude that our choice of limb darkening coefficients does not appear to have a significant impact on our results. We also allowed the limb darkening coefficients to vary as free parameters and found best-fit coefficients $\rm u_{1}=0.57_{-0.40}^{+0.92}$, $\rm u_{2}=0.1_{-1.6}^{+1.0}$. We therefore conclude that the limb-darkening in the Spitzer band is poorly constrained by our data, and fix these coefficients to the model values in our fits.

We fitted the combined astrophysical and instrumental noise models to the \emph{Spitzer} data using an ensemble sampler Markov chain Monte Carlo method implemented in the python package {\ttfamily Emcee} (\citealt{2013PASP..125..306F}).  In addition to the four transit shape parameters listed above, our fit also included a linear slope and allowed the photometric scatter $\sigma$ to vary as a free parameter. We placed a Gaussian prior on $a/R_{\rm *}$ to account for independent constraints on the properties of the host star. The value of this prior was calculated from Kepler's third law using the orbital period measured by \citet{2016ApJ...820...41H} and 
the stellar mass and radius reported in \cite{2017ApJ...836..167D}, with a relative width of 19\% that reflects the uncertainties in these parameters. 
After an initial 1500 step burn-in with 60 walkers, we ran the MCMC for an additional $2.7\times10^5$ steps, 
corresponding to a minimum of $6\times10^4$ times the autocorrelation length for each model parameter.  
Table \ref{ta:lcv} lists our best-fit values and corresponding uncertainties for each model parameter, as well as the stellar parameters that we used. We found that our results of the astrophysical parameters does not change over 1$\sigma$ to the different choices of aperture radius, bin size, and trim duration described in \S\ref{sec:obs}. The transit time $T_c$ changed by less than 1.4$\sigma$ with some choices of the trim duration, as it is more sensitive than other parameters to the structure of the light curve. 

\subsection{K2 Light Curve}    
\label{sec:k2lcv}
After carrying out an initial fit to the {\em Spitzer} data alone, we next set out to incorporate constraints from the {\em K2} light curves.  As discussed in \S\ref{sec:k2obs}, we first corrected for temporal baseline variations in the {\em K2} data by fitting a linear model to 3.4 hours of baseline data before the ingress and after the egress for each of the 29 individual transit events. We then divided each trimmed transit segment by the best-fit linear model. 
Finally, we combined the data into a single phased light curve using the ephemeris calculated from this analysis (Table \ref{ta:lcv}). Because the 30 minute integration time of the {\em K2} data is long relative to the one hour transit duration, we used the ``super-sample'' option in {\ttfamily Batman} to calculate the averaged light curve flux shape. 

In the {\em K2} fit we used the same set of four transit parameters ($R_{\rm p}/R_{\rm *}$, $b$, $a/R_{\rm *}$, and $T_{0}$) as in the {\em Spitzer} fits, but also included the orbital period $P$ as an additional free parameter. In these fits we defined $T_{0}$ as the epoch of the first transit observed by {\em K2} (\citealt{2016ApJS..222...14V}).   

As before, we assumed a circular orbit and calculated the {\em K2} transit model using band-integrated quadratic limb darkening coefficients calculated by \citet{2010A&A...510A..21S}.  We also allowed the noise in the {\em K2} light curve to vary as a free parameter, for a total of six free parameters. As in the {\em Spitzer} fit, we also re-ran our fits with limb-darkening coefficients for models $250$ K hotter and log($g$) 0.5 lower than our nominal model, and again found that it had a negligible effect on our results.  Results from each individual fit are listed in Table \ref{ta:lcv}. We find that in the {\em K2} data the RMS scatter of the residuals is $0.00040$, while in the {\em Spitzer} data binned on the {\em K2} cadence (30 minutes) the scatter is $0.00016$, 40\% lower than the {\em K2} scatter.

\subsection{Spitzer and K2 Joint Analysis}
\label{sec:jointlcv} 
After fitting to the {\em Spitzer} and {\em K2} data separately, we finally carried out a joint fit including both the {\em Spitzer} and {\em K2} light curves.  

In the joint version of the fits, we allowed $R_{\rm p}/R_{\rm *}$ to vary separately in the {\em Spitzer} and {\em K2} bands, and assumed common values for the impact parameter $b$ and semi-major axis ratio $a/R_{\rm *}$.  We also fitted the orbital period $P$ and used the center of transit time $T_{0}$ from the previous {\em K2} observation (\citealt{2016ApJS..222...14V}) to calculate the predicted transit center times for the individual {\em K2} transit events at each step in our fit. This version of the fit includes a total of eight free parameters, including four transit shape parameters (${R_{\rm p}/R_{\rm *}}_{\rm Spitzer}$, ${R_{\rm p}/R_{\rm *}}_{\rm K2}$, $b$, $a/R_{\rm *}$,), the orbital ephemeris ($P$ and $T_{0}$), and two noise parameters ($\sigma_{\rm Spitzer}$, $\sigma_{\rm K2}$). After an initial 2000 steps burn-in with 60 walkers, we ran the MCMC for an additional 80000 steps, corresponding to a minimum of 9000 times the autocorrelation time for each model parameter. Results from all the three versions of the fits are listed in Table \ref{ta:lcv}, and the joint fit results are plotted in Figure \ref{fig:lcv}.

\begin{table*} 
\begin{center}
\caption{Transit Light Curves Analyses Results}
\label{ta:lcv}
\begin{tabular}{cccc}
\hline\hline
 & {\em Spitzer} & {\em K2} & $Joint$ \\
\hline 
$R_{\rm p}/R_{\rm *}$ (\em Spitzer) &$0.0795_{-0.0022}^{+0.0023}$ &...& $0.0802\pm0.0022$ \\ 
$R_{\rm p}/R_{\rm *}$ (K2) &...&$0.0755_{-0.0022}^{+0.0038}$ & $0.0788_{-0.0030}^{+0.0034}$ \\ 
$b$ &$0.66_{-0.22}^{+0.11}$&$0.50_{-0.27}^{+0.19}$& $0.68_{-0.09}^{+0.18}$ \\
$T_c$ [BJD\_TDB] & $2457796.26788_{-0.00046}^{+0.00047}$ & $2456980.2503_{-0.00037}^{+0.00039} $& $2457796.26865_{-0.00048}^{+0.00049}$\\
$P$ [days] &...& $2.260449\pm 0.000023$ & $2.2604380\pm0.0000015$\\
$a/R_{\rm *}$ &$14.4_{-2.1}^{+2.6}$&$17.0_{-2.7}^{+2.1}$&$14.4_{-2.5}^{+1.7}$ \\ 
$\sigma_{\rm Spitzer}$ &$(151.7_{-8.1}^{+8.8})\times 10^{-5}$&...& $(150.7_{-8.7}^{+7.9})\times 10^{-5}$\\ 
$\sigma_{\rm K2}$ &...&$(40.2\pm 1.5)\times 10^{-5}$& $(40.2\pm 1.5)\times 10^{-5}$ \\ 
$R_{\rm p}$$^{\rm a}$ ({\em Spitzer}) [$R_{\Earth}$] &$2.43\pm0.28$ &... & $2.45\pm0.28$ \\ 
$R_{\rm p}$$^{\rm a}$ ({\em K2}) [$R_{\Earth}$] &... & $2.30_{-0.26}^{+0.28}$ & $2.40\pm0.28$\\ 
$a$$^{\rm a}$ [AU] & $0.0187_{-0.0034}^{+0.0040}$ & $0.0221_{-0.0043}^{+0.0037}$ & $0.0187_{-0.0039}^{+0.0030}$ \\ 
\hline 
($u_1, u_2$) (\em Spitzer) (Fixed) & $(0.0092, 0.1877)$ & $...$ & $(0.0092, 0.1877)$ \\ 
($u_1, u_2$) (\em K2) (Fixed) & $...$ & $(0.4572, 0.2876)$ & (0.4572, 0.2876) \\ 
\hline\hline
\footnotetext{Stellar parameters: $R_{*} = 0.280\pm0.031 R_{\Sun}$, $M_{*} = 0.201\pm0.096 M_{\Sun}$, $T_{\rm eff} = 3290\pm90$ K, log($g$) = $4.93\pm0.24$, [M/H] = $0.208\pm0.090$, and [Fe/H] = $0.332\pm0.096$ (\citealt{2017ApJ...836..167D})}
\end{tabular}
\end{center}
\end{table*} 

\begin{figure*}
  \centering
  \includegraphics[width=1.0\textwidth]{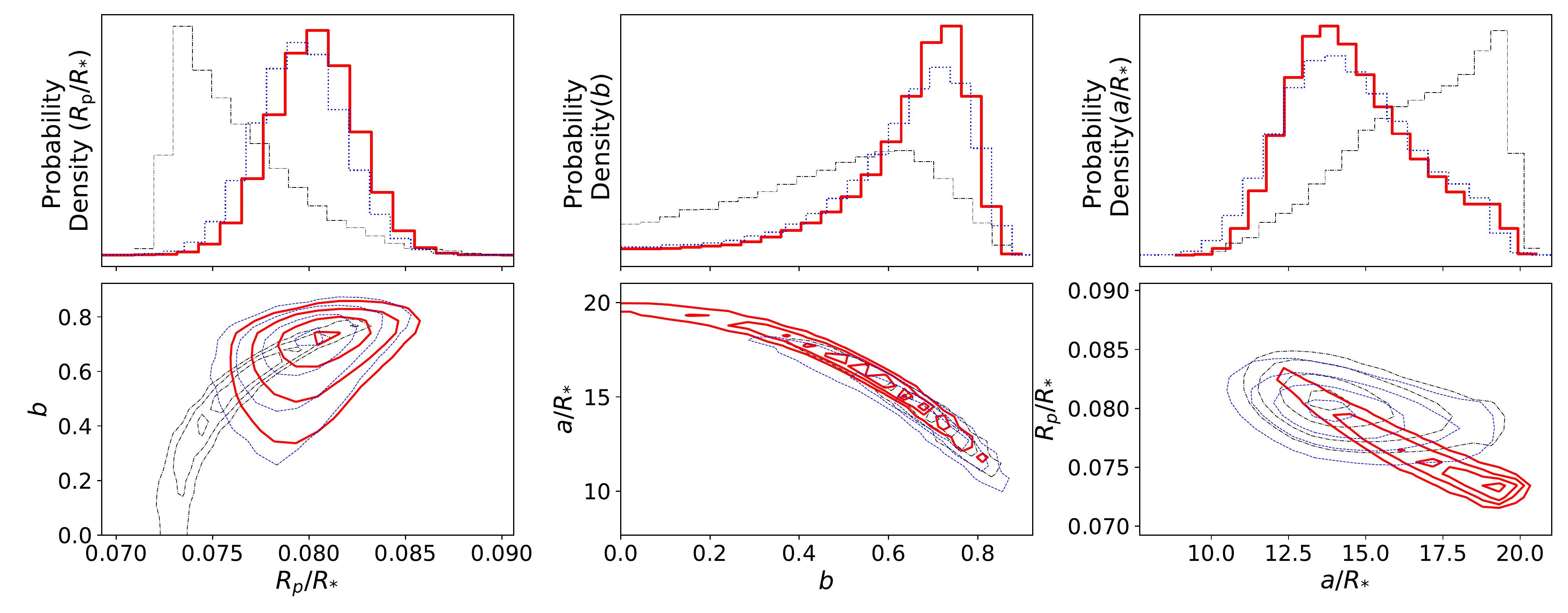}
  \caption{Top plots: normalized 1D posterior probability density functions of parameters $R_{\rm p}/R_{\rm *}$, $b$ and $a/R_{*}$. Bottom plots: 2D probability contours for these same three parameters, with lines indicating the 1, 2, and 3$\sigma$ limits. Results from the joint fit are shown as solid red lines, the {\em Spitzer}-only fit as dashed blue line, and the {\em K2}-only fit as dash-dotted black lines. Here, $R_{\rm p}/R_{\rm *}$ ({\em Spitzer}) is shown for the {\em Spitzer}-only results and the joint results, and $R_{\rm p}/R_{\rm *}$ ({\em K2}) is shown for the {\em K2}-only results. 
\label{fig:pdf}
}
\end{figure*}

\section{Discussion}
\label{sec:disc} 
 
Our new fits improve upon the previous {\em K2} results in two key areas. First, the significantly shorter integration times of our {\em Spitzer} observations and the reduced effects of stellar limb-darkening at long wavelengths allowed us to better resolve the shape of ingress and egress, resulting in smaller uncertainties on $R_{\rm p}/R_{*}$ and $b$, modestly reduced the uncertainties on $a/R_{*}$,  and decreased the degeneracy between $R_{\rm p}/R_{\rm *}$ and the other two parameters. Figure \ref{fig:pdf} shows a comparison of the 1D and 2D posterior probability density functions for these three parameters from the joint fit, the {\em Spitzer}-only fit, and the {\em K2}-only fit. As shown in Table \ref{ta:lcv}, the parameters $b$, $P$, and $a/R_{*}$ obtained from individual fits to each data set as well as the joint fit are consistent to within 1$\sigma$. Our values for $R_{\rm p}/R_{\rm *}$ measured in the {\em Spitzer} and the {\em K2} bands also agree to within 1.5$\sigma$ ($0.0047\pm0.0031$) in all cases. This agreement among different wavelength provides a further evidence that the transit signal is not a false positive due to a blended stellar eclipsing binary (Desert et al. 2015, Hirano et al. 2016). We also set $R_{\rm p}/R_{\rm *}$ equal in both bands in a joint fit and found $R_{\rm p}/R_{\rm *} = 0.0800\pm0.0021$, which is consistent with the results in each individual band. We also confirmed that assuming a common value for $R_{\rm p}/R_{\rm *}$ between the two bands did not change other transit shape parameters by more than 0.5$\sigma$.  Second, the longer baseline (2.24 years) of our new observations results in an improved orbital ephemeris. Our joint fit reduces the uncertainty on the orbital period by more than an order of magnitude as compared to previous results (\citealt{2016ApJ...820...41H}, \citealt{2016ApJS..222...14V}). \citealt{2016ApJ...820...41H} reported two sets of ephemerides, one from {\em K2} only, and another jointly fitting the {\em K2} observation as well as ground-based transit observations made nine months later. The K2-only results predict a transit time of our {\em Spitzer} observation that is $1.7\pm1.0$ hours earlier than our measurement. The inclusion of ground-based transit photometry improves the accuracy and precision, giving a predicted transit time that is $10\pm21$ minutes later than our measured time. We note that our fits utilize an improved version of the {\em K2} photometry provided by Andrew Vanderburg (\citealt{2016ApJS..222...14V}), which appears to perform better than the version of the {\em K2} photometry used in \citet{2016ApJ...820...41H}. Since \citealt{2016ApJS..222...14V} didn't estimate uncertainties on transit parameters, we use the MCMC uncertainties from our K2-only fit as a proxy. Using their ephemeris, the prediction is only $4\pm12$ minutes later than our measured time, so the precision improves by a factor of five. We conclude that the quality of the {\em K2} photometry can make a significance difference in the precision of the original orbital ephemerides (see \citealt{2017ApJ...834..187B} for another example), but this difference is largely erased by the addition of follow-up transit observations with a significantly longer baseline.

Our newly improved ephemeris will be essential for efficient scheduling of observations with {\em JWST}. If we consider a hypothetical transit observation in mid-2019, and if only the K2 results reported in \citealt{2016ApJ...820...41H} were available, there would be an 2.2-hour uncertainty in the transit time. A reasonably conservative observing plan would be to cover $\pm2\sigma$ in the predicted transit time, requiring 10 hours of {\em JWST} time to observe a one-hour transit event. Even more worrying, the predicted time from the {\em K2}-only result would be 4.6 hours earlier than our new updated estimate, indicating that the actual transit would be centered just 0.2 hours after the start of the observations (i.e., we would likely have missed ingress).  If we instead utilize our new ephemeris, the uncertainty on the predicted transit time is only 1.5 minutes, and the corresponding $\pm2\sigma$ window is reduced to $\sim$one hours, allowing us to reliably center the transit in a $2-3$ hour observing window.

Although the inclusion of the new {\em Spitzer} transit observation allowed us to reduce the uncertainty in the planet's orbital period by more than an order of magnitude, our updated estimates for the planet's properties also benefited from the use of refined stellar parameters from \citet{2017ApJ...836..167D}. 

Using these refined stellar parameters and our joint fit to the {\em K2} and {\em Spitzer} light curves, we find that the planet has a radius of $2.45\pm0.28 \rm R_{\Earth}$ in the 4.5 $\mu$m {\em Spitzer} band and $2.40\pm0.28 \rm R_{\Earth}$ in the {\em K2} band, and an effective temperature of $590\pm60$K, assuming an albedo of $0.15\pm0.15$. These results agree with \citet{2016ApJ...820...41H}'s value of $2.32\pm0.24 \rm R_{\Earth}$ within 1$\sigma$, although we prefer a modestly larger planet radius.  

As first pointed out by \citet{2016ApJ...820...41H}, K2$-$28b is a particularly favorable target for transmission spectroscopy of small and cool planets. We evaluate the uniqueness of this target by calculating the transmission and secondary eclipse signals for all of the currently confirmed planets listed in the NASA Exoplanet Archive with radii between $1-3$ $R_{\Earth}$ orbiting M dwarfs with magnitudes less than 12 in $K$-band, resulting in a comparison sample of 34 planets.  For transmission spectroscopy, we assume that the atmosphere has a thickness of 2 scale heights (\citealt{2017AJ....154..261C}) and assume a hydrogen-dominated composition with a mean molecular weight of $\rm H_2$. For planets without measured masses, we use the probabilistic mass-radius relation from \citet{2016ApJ...825...19W} 
to estimate masses using the observed radii, and assume an albedo of $0.15\pm0.15$ when calculating their equilibrium temperatures. Our estimates show that K2$-$28b has a scale height of $\sim$0.01 of its radius, and produces a transmission signal of $\sim$200 ppm, making it the seventh best candidate for transmission spectroscopy among the 34 selected planets.

\begin{figure}[]
  \centering
  \includegraphics[width=0.5\textwidth]{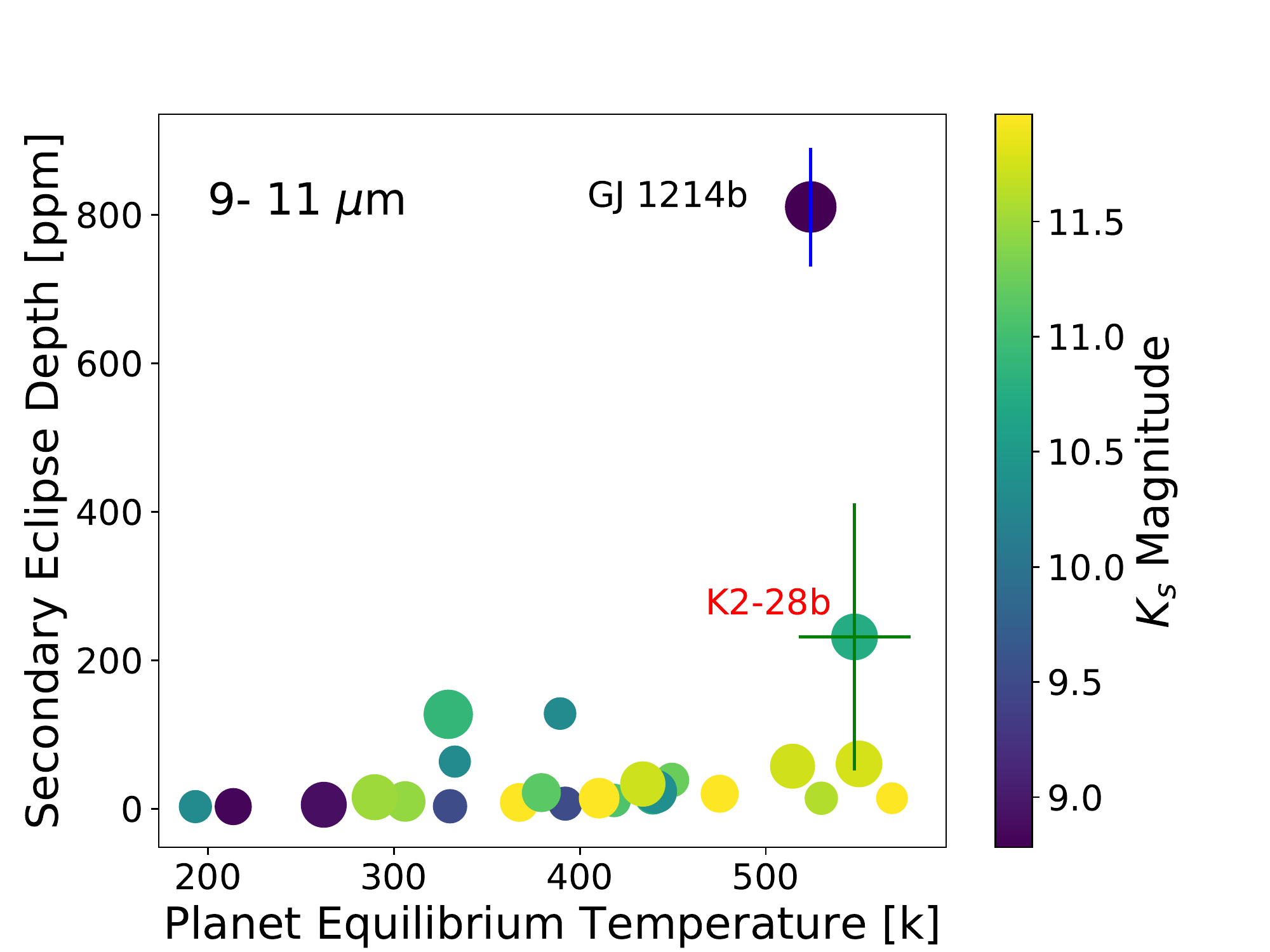}
  \caption{Estimated secondary eclipse depth vs. planet equilibrium temperature in wavelength 9-- 11 $\mu$m for small ($R_{\Earth} < R_{\rm P} < 3R_{\Earth}$) and cool ($T_{\rm eq}<600$ K) planets orbiting bright ($K_{\rm mag} < 12$) M-dwarfs. $K$ magnitudes for each star are indicated by the color bar on the right. Points sizes represent planet radius. The error bars of GJ 1214b and K2$-$28b were estimated using the {\em JWST} ETC with the Mid-Infrared Instrument (MIRI) imaging mode in the 10 ${\mu}$m band. Note that they are the measurement uncertainties, not the astrophysical uncertainties. 
\label{fig:sec_ecl} 
}
\end{figure}

\begin{figure}[]
  \centering
  \includegraphics[width=0.5\textwidth]{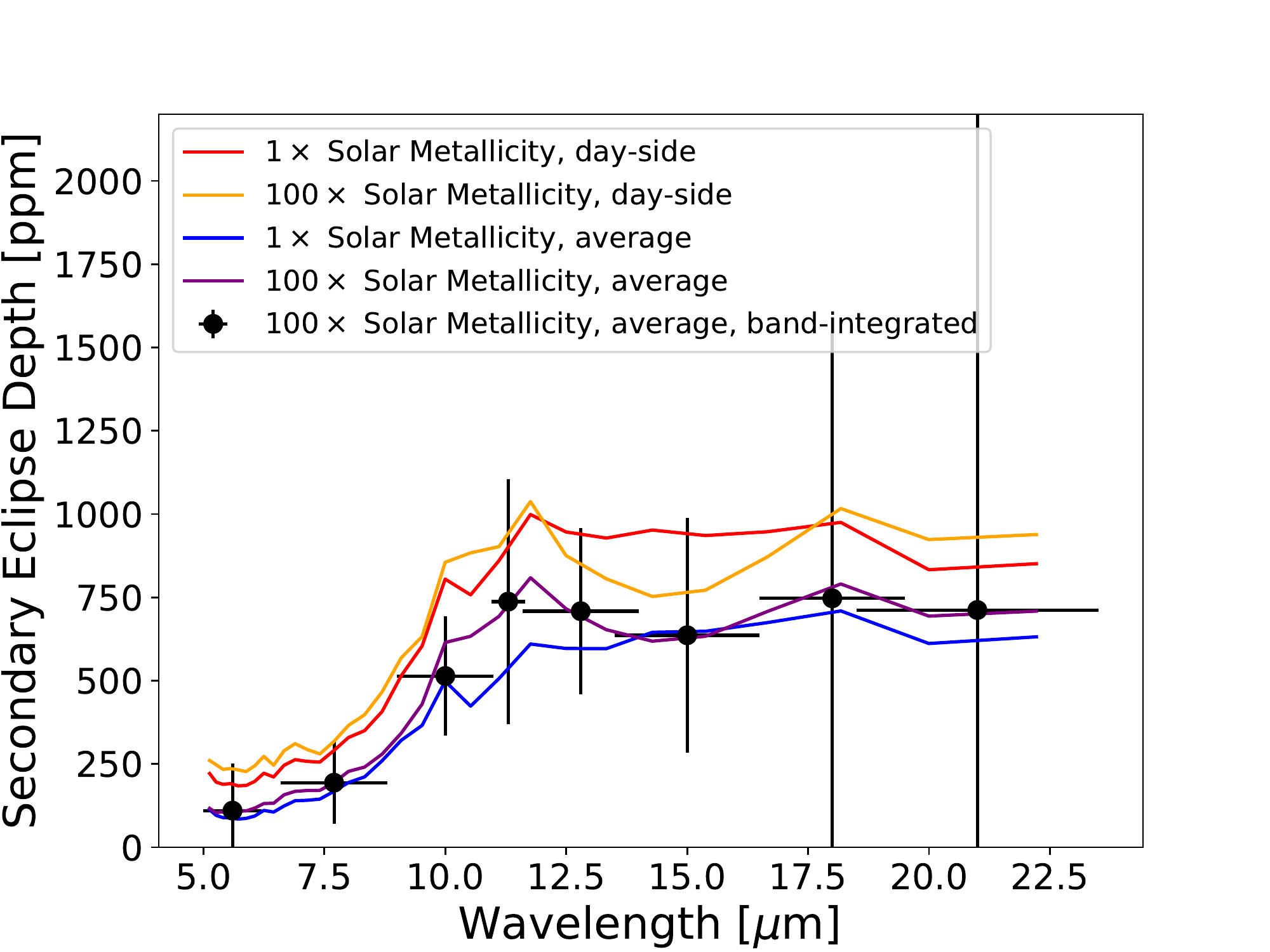}
  \caption{Predicted dayside emission spectra for K2$-$28b calculated assuming solar C/O ratios, equilibrium chemistry,
and no clouds. We consider both the $1 \times$ and $100\times$ solar metallicity cases, as well as scenarios with either minimal or full recirculation between the day and night side hemispheres.  We show the band-integrated values for the $100 \times$ solar efficient recirculation case as filled black circles, and estimate the corresponding uncertainties in each band for a single eclipse observation using the online Exposure Time Calculator for JWST's MIRI instrument. 
\label{fig:real_model} 
}
\end{figure}

Due to the apparent prevalence of clouds in the atmospheres of smaller, cooler transiting planets (e.g., \citealt{2017arXiv170800016C}), secondary eclipse observations offer an important and complementary window into the atmospheres of these planets (e.g., \citealt{2015ApJ...815..110M}; \citealt{2017AJ....153...86M}).  We select 26 planets predicted to be colder than 600 K from the previous sample of 34 planets, and calculate their predicted secondary eclipse depths assuming blackbody spectra for both the star and the planet.  As before, we assume an albedo of $0.15\pm0.15$ when calculating the planet's equilibrium temperature. As shown in Figure \ref{fig:sec_ecl}, K2$-$28b has the second-deepest eclipse depth in this sample after GJ 1214b, but orbits a significantly fainter host star.  We explore the favorability of these two planets for secondary eclipse observations with {\em JWST} by calculating the predicted SNR for single eclipse observations with the Mid-Infrared Instrument (MIRI) imaging mode in the 10 ${\mu}$m band using the {\em JWST} ETC\footnote{https://jwst.etc.stsci.edu}.  We note that GJ 1214b has been selected by the MIRI instrument team as a {\em JWST} Cycle 1 target in exactly this mode, which is ideal for detecting secondary eclipses of small, cool transiting planets. We find a predicted secondary eclipse signal of $230\pm180$ ppm for MIRI 10 $\mu$m observations of K2$-$28b, while GJ 1214b has a more favorable prediction of  $700\pm80$ ppm in this band. However, we note that the presence of molecular features in the planet's dayside emission spectrum could increase the brightness temperature relative to the blackbody prediction by a factor of $2-3$ in some bands, perhaps making this planet detectible with a relatively modest number of eclipse observations.  

We explore this scenario in more depth using models of the thermal emission spectrum of K2$-$28b. These models are calculated assuming radiative-convective and chemical equilibrium using methods described in more detail in \citet{{2008ApJ...678.1419F}, {2017ApJ...850..121M}}. All models assume solar C/O ratios, equilibrium chemistry, and are cloud-free. Figure \ref{fig:real_model} shows the resulting dayside emission spectra for K2$-$28b assuming different atmospheric metallicities and heat redistribution efficiencies. We focus on the $100\times$ solar metallicity case with efficient recirculation, as planets of this size appear to have relatively large core-mass fractions (e.g., \citealt{2015ApJ...801...41R}) and planets cooler than approximately 1000 K have relatively efficient day-night circulation patterns (e.g., \citealt{2015ApJ...810..118K}, \citealt{2015MNRAS.449.4192S}, \citealt{2017ApJ...850..154S}). In this scenario we find that the signal-to-noise ratio for a single eclipse observation is highest in the MIRI 10 $\mu$m band, with a predicted secondary eclipse signal of $450\pm180$ ppm.  This planet is currently the only small ($< 3 R_{\Earth}$) and cool ($<$ 600 K) planet aside from GJ 1214b with a potentially detectable secondary eclipse.

\begin{figure}
  \centering
  \includegraphics[width=0.5\textwidth]{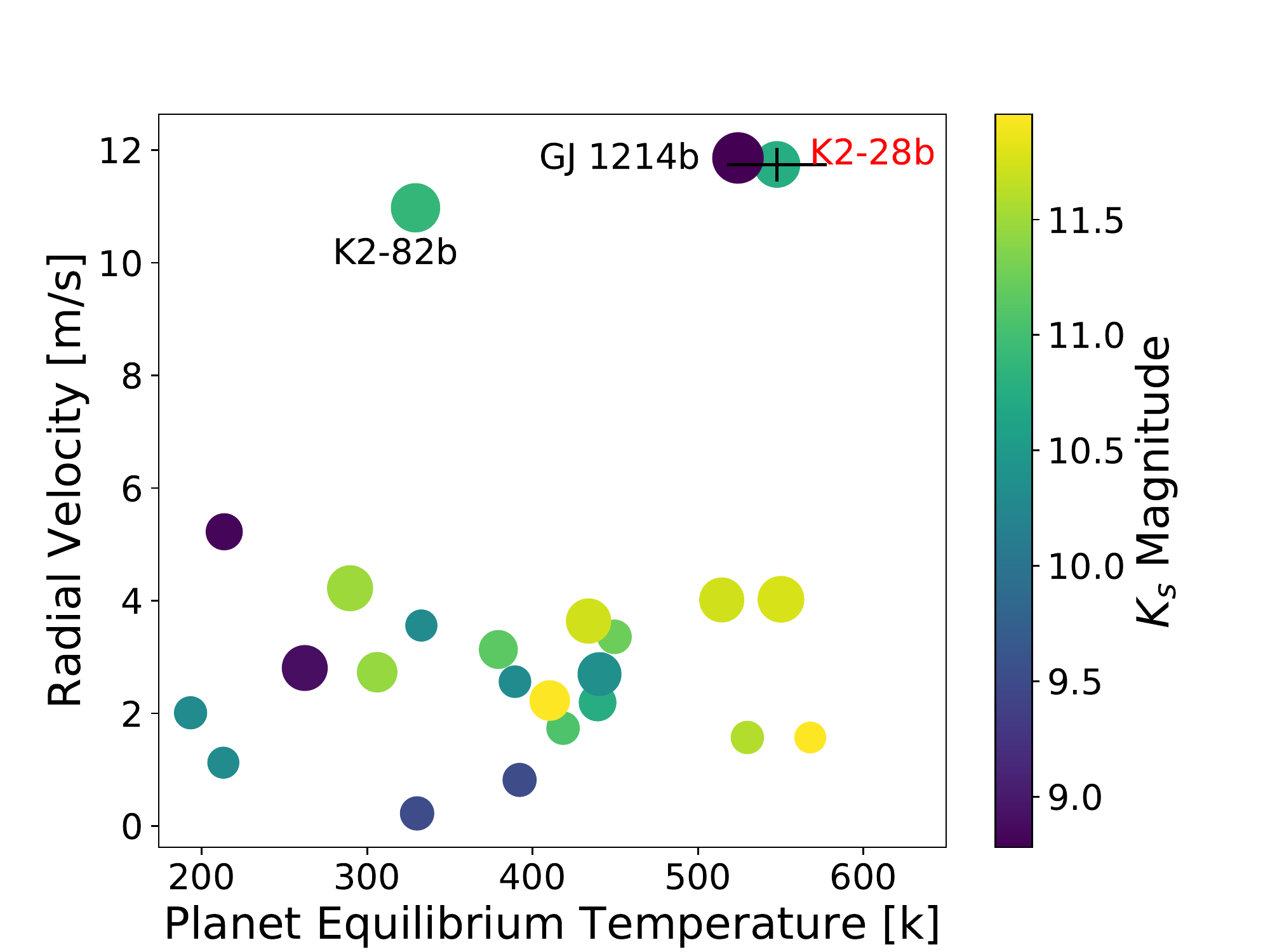}
  \caption{Estimated radial velocity vs. planet equilibrium temperature for small ($R_{\Earth} < R_{\rm P} < 3R_{\Earth}$) and cool ($T_{\rm eq}<600$ K) planets orbiting bright ($K_{\rm mag} < 12$) M-dwarfs. $K$ magnitudes for each star are indicated by the color bar on the right. Points sizes represent planet radius. 
\label{fig:RV} 
}
\end{figure}

K2$-$28 is a mid-M dwarf with only one planet detected. It has been suggested that planet formation around mid-M dwarfs might be more efficient than around solar-type stars, as $21^{+7}_{-5}$ \% of them host multiple planets with periods of less than 10 days (\citealt{2015ApJ...801...18M}). Although there do not appear to be other transiting planets in the K2-28 system, we can search for evidence of additional non-transiting planets using radial velocity observations. K2$-$28 is a good target for upcoming red-optical and near-infrared RV measurements, as its spectrum peaks at $\sim0.9\mu$m and K2-28b is expected to induce a relatively large radial velocity signal. We determine the relative favorability of K2-28b for radial velocity follow-up by comparing its predicted RV semi-amplitude to that of the other 26 planets with radii between 1-3 $R_{\Earth}$ and predicted temperatures less than 600 K orbiting M dwarfs with magnitudes less than 12 in the K-band. For planets without published mass measurements, we estimate predicted masses using the probabilistic mass-radius relation from \citet{2016ApJ...825...19W}. For K2$-$28b this relation predicts a mass of $8\pm2 M_{\Earth}$, corresponding to a RV semi-amplitude of $11.7\pm4.0$ m/s.  As shown in Figure \ref{fig:RV}, K2$-$28b is one of the only three planets that produce radial velocities greater than 10 m/s, and is the smallest among them. Due to its relatively low brightness, K2$-$28 is best observed by upcoming RV instruments on large telescopes, such as the Habitable Planet Finder on the Hobby-Eberly Telescope, and the Infrared Doppler instrument for the Subaru Telescope, whose design precisions are between 1-3 m/s (\citealt{2012SPIE.8446E..1SM}, \citealt{2014SPIE.9147E..14K}). We estimate an expected measurement error of 0.3 m/s for K2-28b (Fig.\ref{fig:RV}) assuming 20 measurements with a measurement error of 1 m/s randomly distributed in orbital phase.  We therefore conclude that this signal should be easily detectable for instruments that can achieve a precision of a few m/s for this star.

One of the primary motivation for study planets around mid-to-late-M dwarfs is to allow access to a population of small and cool planets that will not be otherwise characterizable around hotter stars. Though {\em K2} has surveyed over 300,000 stars up to date, K2$-$28b remains one of only eight K2 planets around mid-to-late-M stars. Among those systems, K2$-$28b is the smallest and coolest planet orbiting a relatively bright host star ($K_{\rm mag}<11$), and it is the brightest host star with $\rm m_{*} < 0.25$ M$_{\Sun}$ in this band. In addition to considering the sample of confirmed transiting planets by current surveys, we also compare K2$-$28b with a simulated catalog of planets from the upcoming {\em TESS} survey described in \cite{2015ApJ...809...77S}. Among the 1984 simulated planet detections, 73\% have radii between 1-3 $R_{\Earth}$; if we select only the subset with predicted equilibrium temperatures cooler than 590 K and magnitudes brighter than 10.75, we find that {\em TESS} should detect 565 relatively bright stars with small and cool planets orbiting. Among these 565 stars, 342 are M-dwarfs, and $\sim$73 are mid-to-late M-dwarfs like K2$-$28.

K2$-$28 is a metal-rich star with a metallicity ([Fe/H]) of $0.332\pm0.096$ (\citealt{2017ApJ...836..167D}). It has been previously noted that mid-M dwarfs hosting planets with radius $R_{\rm P} > 2R_{\Earth}$ tend to be more metal-rich (\citealt{1538-3881-155-3-127}). We independently reproduce this calculation using our revised radius measurement for K2$-$28b, and utilizing all of the confirmed planets orbiting M dwarf later than M3 identified in the NASA Exoplanet Archive. After excluding systems without [Fe/H] measurements and uncertainties, we obtain a sample of 15 stars and 28 planets. We find that stars hosting planets with $R_{\rm P} \leq 2R_{\Earth}$ have an average metallicity of $-0.103\pm0.001$, while stars hosting planets with $R_{\rm P} > 2R_{\Earth}$ have a metallicity of $0.104\pm0.015$, 11$\sigma$ greater than the former group.  We therefore conclude that there continues to be strong evidence for this correlation. Note that there may be detection biases in this sample. Metal-rich stars are bigger (see Fig 23 of \citealt{0004-637X-804-1-64}), so it's harder to find small planets orbiting metal-rich stars (e.g., \citealt{2014Natur.509..593B}). The correlation may be related to formation timescales.  In this picture, metal-rich disks form more massive cores earlier on, allowing the growing cores to accrete a small amount of hydrogen gas.  

With a radius of $2.45\pm0.28 \rm R_{\Earth}$, K2-28b is very likely to host a hydrogen-rich atmosphere with a mass equal to a few percent of the planet's total mass (\citealt{2014ApJ...792....1L}, \citealt{2015ApJ...801...41R}), increasing its favorability for transmission spectroscopy (\citealt{2009ApJ...690.1056M}).  This radius is also consistent with the picture described in \citet{2017AJ....154..109F} and \citet{2017ApJ...847...29O}, among others, in which more highly irradiated planets tend to lose their hydrogen-rich atmospheres while those at lower irradiation levels are able to retain them, resulting in a bimodal radius distribution for close-in planets. K2$-$28b's radius places it firmly on the large-radius side of this bimodal distribution, as would be expected for a relatively cool planet.  K2$-$28b is an ideal system for testing this hypothesis, as it is favorable for both infrared radial velocity mass measurements and atmospheric characterization with {\em JWST}. 

\section{Conclusion}
\label{sec:con}
In this paper we present a new 4.5 $\micron$ Spitzer transit observation of the sub-Neptune K2$-$28b, which we combine with previously published optical photometry from K2. We carry out a simultaneous fit to both data sets in order to derive improved estimates for the planet's radius, semi-major axis and orbital ephemeris, allowing us to predict transit times for this planet with a precision of 1.5 minutes (1 sigma) during {\em JWST} 's first year of observation. We also consider the relative favorability of this planet for atmospheric characterization with {\em JWST} as compared to the sample of other small ($< 3R_{\Earth}$) and cool ($<600$ K) transiting planets confirmed to date.  We find that K2$-$28b and GJ 1214b are the only two planets in this sample with potentially detectible secondary eclipses, and use atmosphere models for K2$-$28b to estimate its secondary eclipse depth in the MIRI bands (5-22 \micron).  We also evaluate the suitability of this planet for radial velocity follow-up, and conclude that it should be detectible with red-optical or near-infrared instruments on larger telescopes. In addition, we compare our results with the simulated {\em TESS} catalog and find K2$-$28b to be typical of the mid-to-late M systems that should be detectable in the {\em TESS} sample. Finally, we find strengthened evidence for a previously reported correlation that mid-M dwarfs hosting planets with radius $R_{p} > R_{\Earth}$ tend to be more mental-rich, based on a tripled sample size and our new results. 

\medskip


This work was performed in part under contract with the Jet Propulsion Laboratory (JPL) funded by NASA through the Sagan Fellowship Program executed by the NASA Exoplanet Science Institute. H.A.K. and C.D.D acknowledge support from the Sloan Foundation and from NASA through an award issued by JPL/Caltech. This work is also based in part on observations made with the Spitzer Space Telescope, which is operated by the Jet Propulsion Laboratory, California Institute of Technology under a contract with NASA. 

\bibliography{K2_28bib}
\bibliographystyle{apj}


\end{document}